\documentstyle[floats,aps,aipbook,epsf]{revtex}

   \def\unlock{\catcode`@=11}

   \unlock

   \def\gsim{\mathrel{\mathpalette\@versim>}}

   \def\lsim{\mathrel{\mathpalette\@versim<}}

   \def\@versim#1#2{\vcenter{\offinterlineskip
        \ialign{$\m@th#1\hfil##\hfil$\crcr#2\crcr\sim\crcr } }}

\newcommand{\zp}[3]{Z. Phys.\ {\bf C#1}, #2 (19#3)}
\newcommand{\plb}[3]{Phys.\ Lett.\ {\bf #1B}, #2 (19#3)}
\newcommand{\np}[3]{Nucl.\ Phys.\ {\bf B#1}, #2 (19#3)}
\newcommand{\prevd}[3]{Phys.\ Rev.\ {\bf D#1}, #2 (19#3)}
\newcommand{\prevl}[3]{Phys.\ Rev.\ Lett.\ {\bf #1}, #2 (19#3)}

\newcommand{\prep}[3]{Phys.\ Rep.\ {\bf C#1}, #2 (19#3)}

\newcommand{\beq}{\begin{equation}}
\newcommand{\eeq}{\end{equation}}
\newcommand{\bea}{\begin{eqnarray}}
\newcommand{\eea}{\end{eqnarray}}

\newcommand{\ra}{\rightarrow}

\newcommand{\nn}{\nonumber}
\renewcommand{\a}{\alpha}
\newcommand{\s}{\sigma}
\newcommand{\hsig}{\hat \sigma}
\newcommand{\ptp}{p_\perp}

\begin{document}

\title{Theory of Hard Diffraction and Rapidity Gaps}

\author{Vittorio Del Duca\thanks{Talk presented at the $10^{th}$ Topical
Workshop on Proton-Antiproton Collider Physics, Fermilab, Batavia, Illinois,
May 9-13, 1995.}}

\address{Deutsches Elektronen-Synchrotron \\
DESY, D-22603 Hamburg , GERMANY}

\maketitle
\begin{abstract}
In this talk we review the models describing the hard diffractive
production of jets or more generally high-mass states in presence of
rapidity gaps in hadron-hadron and lepton-hadron collisions.
By rapidity gaps we mean regions on the
lego plot in (pseudo)-rapidity and azimuthal angle where no hadrons are
produced, between the jet(s) and an elastically scattered hadron (single
hard diffraction) or between two jets (double hard diffraction).
\end{abstract}

Single hard diffraction has been observed by the UA8 Collaboration
\cite{ua8,shard} at the CERN S$p\bar{p}$S Collider ($\sqrt{s} = 630$ GeV),
and by the H1 and Zeus Collaborations \cite{zeus,fdif,fdiz,ng}, in deep
inelastic scattering events (DIS) at the DESY HERA
$e\, p$ Collider ($\sqrt{s} = 296$ GeV). Double hard diffraction has been
observed by the CDF and D0 Collaborations \cite{gaps,bert} at the Fermilab
Tevatron
$p\,\bar{p}$ Collider ($\sqrt{s} = 1.8$ TeV), and in photoproduction events by
the Zeus Collaboration at HERA \cite{laura}. The distinguishing feature between
single and double hard diffraction is the momentum transfer $t$: while
$|t|\simeq 1-2\,\rm{GeV}^2$ in the UA8 experiment, and
$|t|\lsim \rm{a\, few\, GeV}^2$ in the DIS events
at HERA, it is very large, $|t|\gsim 10^3\,\rm{GeV}^2$, in the Tevatron
experiments, which suggests that short-distance strong-interaction physics
must play a fundamental role in the latter.

\section{Single Hard Diffraction}

\subsection{The Ingelman-Schlein model}
\label{sec:one}

\begin{figure}[htb]
\vskip-0.5cm
\epsfysize=5.5cm
\centerline{\epsffile{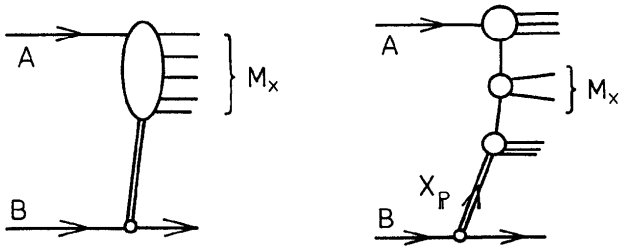}}
\vskip-1cm
\caption{($a$) Single diffraction, and ($b$) with jet production, in
high-energy hadron-hadron collisions.}
\label{fig:one}
\end{figure}

Diffractive production of jets was predicted by Ingelman and Schlein (IS) to
occur in hadron-hadron collisions at high energies \cite{is}. In order to
understand the IS model, let us consider ordinary single diffraction at
high c.m. energies $\sqrt{s}$, i.e. a collision between two hadrons,
$A$ and $B$, where hadron $B$ is elastically scattered and hadron $A$
fragments into a high-mass $M_X$ state, $M_X^2\gsim 10\,\rm{GeV}^2$ and $(M_X^2
/s)\lsim 0.1$, with a gap in hadron production between hadron $B$ and the
fragments of hadron $A$ (Fig.~\ref{fig:one}a). This process is
phenomenologically well described by Regge theory \cite{pdb} through the
exchange of a colorless object, conventionally termed {\sl pomeron},
which accounts for the gap,
\beq
{d\sigma\over dt\, dx_{I\!\! P}} = f_{B I\!\! P}(x_{I\!\! P},t)\,
\sigma_{tot}(A I\!\! P)\,
,\label{one}
\eeq
where $t$ is the momentum transfer and $x_{I\!\! P} = M_X^2/s$. The pomeron
flux factor, $f_{B I\!\! P}$, describes the emission of the pomeron from
hadron $B$, and the cross section, $\sigma_{tot}(A I\!\! P)$, describes
the scattering between the pomeron and hadron $A$ with creation of a
high-mass state,
and is given in terms of a triple pomeron coupling, which is intuitively
apparent when we consider the square of the diagram of Fig.~\ref{fig:one}a.
The gap width must satisfy the kinematic constraint
$\Delta\eta_{gap} \gsim \ln(1/x_{I\!\! P})$ (cf. sec.\ref{sec:due}).

Regge theory does not say what the pomeron is. There are however
perturbative models where the pomeron is pictured as a colorless two-gluon
bound state \cite{ln,fkl}. Ingelman and Schlein
proposed that if the pomeron had a partonic substructure it should manifest
itself in the high-mass diffractive scattering through the appearance of jets
(or heavy quarks \cite{bcss}). Inclusive jet production in
hadron-hadron collisions is described by the factorization formula,
\beq
d\sigma(A + B \ra jet(s) + X) = \sum_{ab} \int dx_a dx_b f_{a/A}
(x_a,\mu) f_{b/B}(x_b,\mu) d\hsig_{ab}\, ,\label{due}
\eeq
where $x_{a(b)}$ is the momentum fraction of parton $a(b)$ within hadron
$A(B)$, $\mu$ is the factorization scale of the order of the jet transverse
energy $E_{\perp}$ and $d\hsig_{ab}$ is the jet production rate at the
partonic level. Substituting the pomeron-proton cross section $\sigma_{tot}(A
I\!\! P)$ in eq.(\ref{one}) with the inclusive jet production rate,
eq.(\ref{due}), we obtain the diffractive jet production rate
(Fig.~\ref{fig:one}b),
\beq
{d\sigma\over dt\, dx_{I\!\! P}} = f_{B I\!\! P}(x_{I\!\! P},t)\, \sum_{ab}
\int dx_a d{x_b\over x_{I\!\! P}} f_{a/A}(x_a,\mu) f_{b/I\!\! P}\left(
{x_b\over x_{I\!\! P}},\mu\right) d\hsig_{ab}\, ,\label{tre}
\eeq
where $x_b/x_{I\!\! P}$ is the momentum fraction of parton $b$ within the
pomeron \cite{is}. We take the pomeron flux factor to be used in eq.(\ref{tre})
as given in ref.~\cite{chpww},
\beq
f_{B I\!\! P}(x_{I\!\! P},t) = {1\over 8\pi^2}\, |\beta_{B I\!\! P}(t)|^2\,
x_{I\!\! P}^{1-2\a(t)}\, ,\label{four}
\eeq
normalized\footnote{Goulianos \cite{dino,din2} argues that the
flux \cite{dl} is not appropriate to describe the high-energy $p\bar{p}$
single-diffraction data of the CERN
UA4 and the Tevatron E710 and CDF Collaborations \cite{sd}.
Interpreting the flux as a probability density of pomerons in the hadron, he
renormalizes it in such a way to never exceed the unity. The ZEUS
Collaboration,
though, claims \cite{fdiz} that the Regge scaling of $F_2^D$ yielded by
the flux of ref.~\cite{dino,din2} does not agree with its data.}
in order to agree with the Donnachie-Landshoff flux factor
\cite{dl} (even though different in its functional form in $t$). The
pomeron-proton coupling, $\beta_{B I\!\! P}(t)$, and the pomeron trajectory,
$\a(t)$, may be obtained from fits to the elastic hadron-hadron cross section
at small $t$ \cite{chpww,dl2},
\bea
\beta_{P I\!\! P}(t) &=& \beta_{\bar{P} I\!\! P}(t) \simeq 4.6\, mb^{1/2}\,
e^{1.9 GeV^{-2} t}\, ,\label{five}\\
\a(t) &\simeq& 1.08 + 0.25\, GeV^{-2}\, t\, .\nn
\eea
Then in order to use eq.(\ref{tre}) one must know the parton densities
in the pomeron. Ingelman and Schlein assumed a pomeron made of gluons and
tested hard, $x f_{g/I\!\! P}(x)=A x(1-x)$, and soft, $x f_{g/I\!\! P}(x)=B
(1-x)^5$, gluon densities in the pomeron, with the constants $A$ and $B$
determined
from the momentum sum rule, $\int_0^1 dx\, x f_{g/I\!\! P}(x) = 1$. However,
since the pomeron is not a particle there is no reason to expect a momentum
sum rule to hold \cite{bcss,dl}.

In addition, eq.(\ref{tre}) entails that the
factorization picture of eq.(\ref{due}), well established in inclusive
processes \cite{css}, carries over to hard diffractive processes, i.e.
there is in the parton density in the proton $f_{p/P}(x,\mu)$ a diffractive
component which factorizes as the whole function $f$ does. However, we
will discuss in sect.~\ref{sec:tre} factorization-breaking effects for which
eq.(\ref{tre}) is violated.

\subsection{Diffractive DIS}
\label{sec:due}

\begin{figure}[htb]
\vskip 0cm
\epsfysize=4cm
\centerline{\epsffile{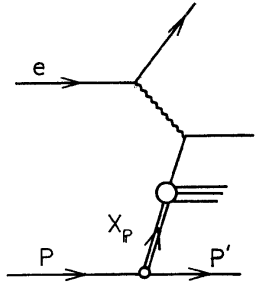}}
\caption{Diffractive DIS diagram $e + p \ra e + p + X$.}
\label{fig:due}
\end{figure}

Let us consider the diffractive deep inelastic scattering (DDIS) $e + p \ra
e + p + X$, where a proton of momentum $P$ is elastically scattered to a
final-state proton of momentum $P'$ (Fig.\ref{fig:due}). The relevant
kinematic invariants are the squared photon-proton c.m. energy,
$W^2 = (q+P)^2$, the momentum transfer $t = (P-P')^2$ and the squared mass
of the hadronic system $M^2 = (q+P-P')^2$. The usual variables of
DIS are $Q^2 = -q^2$ and $x_{bj} = Q^2/(W^2+Q^2-M_P^2)$, in
terms of which we can introduce the variables \cite{fdif,fdiz},
\bea
x_{I\!\! P} &=& {q\cdot (P-P')\over q\cdot P} = {M^2+Q^2-t\over Q^2} x_{bj}\,
,\label{six}\\ \beta &=& {x_{bj}\over x_{I\!\! P}}\, .\nn
\eea
We can parametrize the proton momentum loss in light-cone coordinates as,
\beq
p_s = P-P' = \left(zP^+, {m_P^2\over P^+}-{m_{P'}^2+p_{s\perp}^2\over
(1-z)P^+}, {\bf p_{s\perp}}\right)\, ,\label{seven}
\eeq
with $P^+ = 2P^0$. If $p_s^+ \gg p_s^-, p_{s\perp}$, which for the HERA
lab frame ($P^0 = 820$ GeV) holds as long as $z \gsim 10^{-6}$, we can
rewrite the pomeron momentum as $p_s = z P$, and we readily find that
$z = x_{I\!\! P}$. It is then easy to derive that the invariant mass of the
system recoiling against the proton is $M_{eX}^2 \simeq x_{I\!\! P} s$, and
that the gap width has lower bound $\Delta\eta_{gap}\gsim \ln(1/x_{I\!\! P})$.

$e + p \ra e + X$ DIS is fully inclusive over the final-state hadrons, thus it
may be parametrized in terms of two structure functions,
\beq
{d\s\over dx_{bj} dQ^2} = {4\pi \a^2\over x_{bj} Q^4}\, \left[(1-y) F_2(x_{bj},
Q^2) + x_{bj} y^2 F_1(x_{bj}, Q^2)\right]\, ,\label{otto}
\eeq
where $y$ is the electron energy loss. If the proton is tagged in the final
state, then in order to describe DIS we need two more structure functions,
however in the kinematic region of DDIS they are negligible since the
transverse momentum of the final-state proton is very small \cite{chpww},
thus we can write,
\bea
& & {d\s\over dx_{bj} dQ^2 dx_{I\!\! P} dt} = \label{nove}\\
& & {4\pi \a^2\over x_{bj} Q^4}\, \left[(1-y) {dF_2^D(x_{bj}, Q^2, x_{I\!\! P},
t)\over dx_{I\!\! P} dt} + x_{bj} y^2 {dF_1^D(x_{bj}, Q^2, x_{I\!\! P}, t)\over
dx_{I\!\! P} dt}\right]\, .\nn
\eea
If the factorization picture of eq.(\ref{tre}) is correct for DDIS we obtain,
\beq
{d\s\over dx_{bj} dQ^2 dx_{I\!\! P} dt} = f_{P I\!\! P}(x_{I\!\! P},t)\,
{1\over x_{I\!\! P}} {d\s\over d\beta dQ^2}\, ,\label{ten}
\eeq
with the flux factor as given in eq.(\ref{four}).
$d\s/d\beta dQ^2$ may be expressed in terms of two pomeron structure
functions like in eq.(\ref{otto}), with the parton momentum
fraction in the pomeron, $\beta$, playing now the role of the Bjorken variable
$x_{bj}$. Comparing eq.(\ref{nove}) to eq.(\ref{ten}), we obtain for example
the diffractive structure function $F_2^D$ in terms of the pomeron structure
function $F_2^{I\!\! P}$,
\beq
{dF_2^D(x_{bj}, Q^2, x_{I\!\! P}, t)\over dx_{I\!\! P} dt} = f_{P I\!\! P}
(x_{I\!\! P},t)\, F_2^{I\!\! P}(\beta, Q^2)\, .\label{elev}
\eeq
Eq.(\ref{elev}) states that $F_2^D$ exhibits the Regge scaling dictated by
single hard diffraction in hadron-hadron scattering,
and the Bjorken scaling typical of the usual DIS. The latter entails that
$F_2^D$ has a leading twist behavior, i.e. it scales in $Q^2$ like the
ordinary $F_2$ structure function. These features are presently in agreement
with the H1 \cite{fdif,ng} and the ZEUS \cite{fdiz,ng} data.

\subsection{Is the factorization picture correct?}
\label{sec:tre}

\begin{figure}[htb]
\vskip 0cm
\epsfysize=5.5cm
\centerline{\epsffile{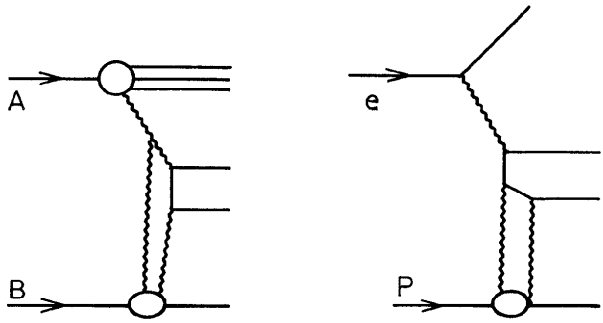}}
\vskip-0.5cm
\caption{Two-gluon exchange ($a$) in hadron-hadron collisions and ($b$) in
DIS.}
\label{fig:four}
\end{figure}

The analysis of sec.\ref{sec:one} and \ref{sec:due} relies upon
the factorization picture of eq.(\ref{tre}). However factorization has not
been proved for any diffractive process.
Following the work of Collins, Frankfurt and Strikman (CFS) \cite{cfs}, we now
illustrate the case of diffractive jet production in hadron-hadron scattering
where there may be non factorizing contributions which spoil the validity
of eq.(\ref{tre}). Let us assume that a pomeron made of two gluons emitted
from hadron $B$ goes wholly into the hard scattering with parton $a$
coming from hadron $A$ (Fig.\ref{fig:four}a). If the two gluon are hard, then
this contribution is suppressed by powers of the scale that characterizes the
hard process (i.e. the jet transverse energy in jet production)
with respect to the non-diffractive contribution due to one-gluon
emission, thus it is higher twist and does not spoil eq.(\ref{tre}).
However if one of the two gluons is soft, it does not contribute to the
power counting in the hard scale and can give a leading-twist contribution.
In the usual factorization where no requirement is made on the final states
such contributions cancel out after summing over all the final-state soft
gluons, however the sum cannot be carried over if one requires a rapidity
gap in the final state because some of the diagrams contributing to the sum
do not form a gap. This can be seen diagrammatically by squaring the diagram of
Fig.\ref{fig:four}a and drawing the cut line which defines the final state in
all the possible positions.

This picture, though, is perturbative and therefore questionable when the
momentum transfer $|t|$ is very small. On these grounds one would expect
the factorization picture of the IS model still to make sense
at very small $|t|$, with non-factorizing contributions growing bigger and
bigger as $|t|$ grows, the signature of these being a leading-twist
contribution with a $\delta(1-\beta)$ dependence on $\beta$.

The UA8 Collaboration \cite{shard}, has examined diffractive jet
production at the S$p\bar{p}$S Collider, with $0.9\,\rm{GeV^2} \le |t|
\le 2.3\,\rm GeV^2$. Assuming the pomeron as made of two gluons
the UA8 Collaboration has found that the data could not be explained invoking
simply a hard gluon density, of the type $\beta f_{g/I\!\! P}(\beta)=A \beta
(1-\beta)$, and that in order to fit the data it was necessary a 30\%
contribution from a $\delta$-function-like component, in agreement with CFS
prediction \cite{cfs}.

In diffractive DIS the two gluons forming the pomeron in Fig.\ref{fig:four}a
cannot couple directly to the photon but must couple to the fermion lines
(Fig.\ref{fig:four}b). In the kinematic region where the perturbative picture
makes sense, i.e. where both the quark and antiquark transverse momenta are of
the order of $Q$, neither of the two gluons is soft and the contribution of
Fig.\ref{fig:four}b is higher twist \cite{chpww,cfs}. Thus the lack of
initial-state interactions suppresses this contribution.

However, it is still possible that the second gluon is soft and is
emitted much later in time as part of the final-state interactions. This
factorization-breaking leading-twist mechanism, conceptually analogous to
the CFS model, has been considered recently by Buchm\"uller and Hebecker
(BH) \cite{bh}, who propose that the rapidity gap
is due to color fluctuations in the long-range final-state interactions
within the proton. BH suppose that the photon-gluon fusion process, which
at the perturbative level accounts for the main contribution to $F_2$ at
small $x_{bj}$ in eq.(\ref{otto}), describes the short-range interaction
also in DDIS (Fig.\ref{fig:four}b). They assume then that the $q\bar{q}$ pair
formed in the hard-scattering process, while propagating in the color field
of the proton, transforms into a color singlet by exchanging a soft gluon with
the proton.

The BH model predicts that $F_2$ and $F_2^D$ have the same Bjorken scaling,
in agreement with the IS model (cf. sec.\ref{sec:due}). As for the Regge
scaling, BH predict that if $F_2 \sim x_{bj}^{-n}$, then
$F_2^D \sim x_{I\!\! P}^{-1-n}$. Thus
in the BH model the Regge scaling is determined by the hard scattering,
and is directly related to the one of the inclusive process. Conversely,
in the IS model the Regge scaling is linked to the pomeron flux factor
(cf. sec.\ref{sec:one}).

As for the CFS model, though, one would expect that the perturbative picture
of the BH model is questionable at very small $|t|$, where the IS model
should still anyway be valid.

\subsection{Parton densities in the pomeron}
\label{sec:four}

If the picture advocated in sec.\ref{sec:one}-\ref{sec:tre} holds it is
possible to fit the data on single hard diffraction to extract the
parton densities in the pomeron. The fits should not be global, i.e.
should not include data from hadron-hadron scattering because of the
factorization-breaking CFS mechanism. However the data from
DDIS and from diffractive direct photoproduction of jets should suffice to
determine the main parton densities. Besides no assumption should be made
on the validity of the momentum sum rule \cite{bcss,dl}. We will follow here
the program proposed in ref.\cite{chpww} to measure the parton densities.

First we note that the pomeron, being an object with the quantum numbers of
the vacuum, has $C = 1$ and is isoscalar. The former property implies that
$f_{q/I\!\! P}(\beta) = f_{\bar{q}/I\!\! P}(\beta)$ for any quark $q$
and the latter that $f_{u/I\!\! P}(\beta) = f_{d/I\!\! P}(\beta)$.
Therefore it is necessary to determine only the up and strange quark densities
and the gluon density. In the parton model the pomeron structure function
in eq.(\ref{elev}) is,
\beq
F_2^{I\!\! P}(\beta, Q^2) = {10\over 9} \beta f_{u/I\!\! P}(\beta,Q^2) +
{2\over 9} \beta f_{s/I\!\! P}(\beta,Q^2) + O(\a_s)\, ,\label{dod}
\eeq
where the gluon density contributes in the $O(\a_s)$ term through the
DGLAP evolution. The gluon density may be directly measured by using data
on jet production from DDIS or diffractive direct photoproduction,
whose rate is,
\bea
& & {d\sigma\over dt\, dx_{I\!\! P}}(e + p \ra e + p + jet_1 + jet_2 + X) =
\label{tred}\\ & & f_{P I\!\! P}(x_{I\!\! P},t) \sum_b \int d{x_b\over
x_{I\!\! P}}\, f_{b/I\!\! P}\left({x_b\over x_{I\!\! P}},\mu\right)
d\hsig(e + b \ra e + jet_1 + jet_2 + X)\, .\nn
\eea
At the lowest order, $O(\a\a_s)$, the final state of the hard scattering,
$jet_1 + jet_2 + X$ consists only of two partons, generated in
quark-exchange and Compton-scattering diagrams for quark-initiated
hard processes, and in photon-gluon fusion diagrams for the gluon-initiated
ones. The parton momentum fraction in the proton, $x_b$, may be
computed from the jet kinematic variables, $x_b = (E/P^0) \exp(2\bar{\eta})$,
with $E$ and $P^0$ the electron and proton energies and $\bar{\eta} =
(\eta_{j_1}+\eta_{j_2})/2$ is the rapidity boost of the jet system. The
pomeron momentum fraction in the proton, $x_{I\!\! P}$ may be obtained, as
noted in sec.\ref{sec:one} and \ref{sec:due}, from the invariant mass of the
system recoiling against the proton, $x_{I\!\! P}\simeq M_{ej_1j_2}^2/s$.

If we neglect the strange quark density these two sets of measurements
suffice to measure the parton densities in the pomeron. From these one can
test if the momentum sum rule, $\int_0^1 dx\, x f_{g/I\!\! P}(x) = 1$,
is correct \cite{msr}. The strange quark
density may then be measured adding to the fit data on charged-current
charm production in DDIS \cite{chpww}.

\subsection{Conclusions}
\label{sec:six}

Single hard diffraction is a well
established phenomenon in hadron-hadron \cite{ua8,shard} and lepton-hadron
\cite{zeus,fdiz} collisions. Several theoretical models have been conceived
which predict or explain these events. They range from models which describe
the strong-interaction process just in perturbative terms \cite{bw,nz}, to
models which rely heavily on soft-interaction modeling and Regge phenomenology
\cite{cap}. We have illustrated the IS model which is a mixture of soft-
and hard-interaction physics.

The IS model yields a consistent description of the HERA data \cite{zeus,fdiz}
on DDIS. The predictions for the Regge and Bjorken scaling for $F_2^D$ are
in agreement with the data at the present level of accuracy. The IS model
relies on the factorization picture, eq.(\ref{tre}), however there may
be factorization-breaking leading-twist contributions due to long-range
initial-state interactions in diffractive hadron-hadron collisions
\cite{shard,cfs}, and due to final-state interactions in DDIS \cite{bh}.
These mechanisms should be more relevant as $t$ grows \cite{chpww,cfs}.
Thus, it is very important to measure the $t$ dependence both in DDIS
and in diffractive jet production at the Tevatron Collider.

In addition, it is questionable that
a momentum sum rule for the parton densities $f_{i/I\!\! P}$ in the pomeron
holds \cite{bcss,dl}, since the pomeron is not an on-shell state.
If we presume that factorization
works only for DDIS and that the momentum sum rule does not apply, we still
have enough information from DDIS and diffractive jet production in direct
photoproduction to fit the parton densities $f_{i/I\!\! P}$ and check the
momentum sum rule \cite{chpww,msr}. In addition, the parton densities
should not depend on $t$ if factorization holds (cf. eq.(\ref{elev})).

The parton densities $f_{i/I\!\! P}$ measured at HERA should be used to
model jet and $W$-boson diffractive production in hadron-hadron collisions
at the Tevatron Collider. If factorization breaks down, the predictions
should disagree with the data.

\section{Double hard diffraction}


\subsection{Gap production in parton scattering}
\label{sec:set}

The initial theoretical motivation for examining two-jet production
with a rapidity\footnote{Here and in the following we identify the true
rapidity used in the theoretical models with the pseudo-rapidity used in
the experiments. There is no difference between the two, as long as we
deal with particles which are massless or for which $\ptp\gg m$, but the
difference must be kept in mind when dealing with the underlying event
\cite{pump}.} gap in hadron production between the jets was $W-W$ boson
scattering via Higgs-boson (or $Z$-boson) exchange at the SSC Collider
\cite{dkt,fad,bj}. Unless the Higgs boson is rather heavy \cite{bp}, $W-W$
boson fusion is not the leading Higgs-boson production mode in hadron-hadron
collisions, the gluon-gluon fusion channel being more important. However
if the final-state $W$'s decay leptonically the $W-W$ boson fusion channel
has the unique signature of a rapidity gap in parton production between
the quarks initiating the scattering. On the contrary, the gluons producing
the Higgs boson would likely radiate more gluons, thus filling the lego plot
with color. Still,
the rapidity-gap signal could be faked by the scattering between color-singlet
two-gluon ladders \cite{bj}, and finally a rapidity gap at the parton
level might not survive the spectator-parton interactions from the underlying
event\footnote{At the LHC Collider, which will operate at high luminosity,
the additional problem of overlapping minimum bias events in a single bunch
crossing arises. However this could be disposed of by requiring a gap in
minijet production rather than in soft-hadron production \cite{bpz}.}.

\begin{figure}[htb]
\vskip 0cm
\epsfysize=5.5cm
\centerline{\epsffile{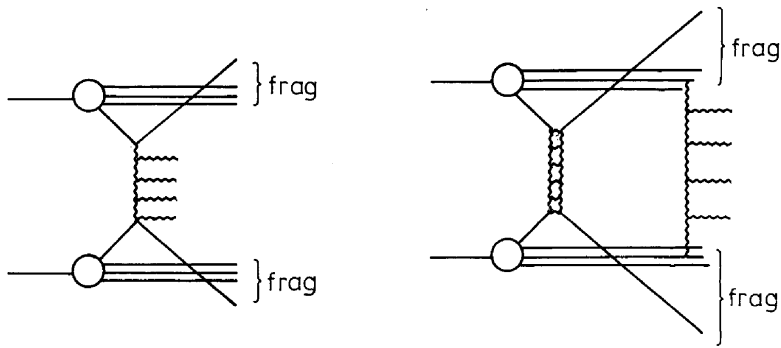}}
\caption{$(a)$ One-gluon and $(b)$ two-gluon exchange in the $\hat t$ channel.
In $(b)$ also the soft-gluon emission from the rescattering between the
spectator partons has been included.}
\label{fig:c}
\end{figure}

The simplest case in which one can start addressing these issues is
two-jet production in hadron-hadron collisions with a rapidity gap in
hadron production between the jets \cite{bj}.
In parton-parton scattering the leading-order process, which is $O(\a_s^2)$
and we may picture through one-gluon exchange in the $\hat t$ channel,
is likely not to produce a gap because the exchanged gluon being a color
octet radiates off more gluons (Fig.\ref{fig:c}a). However a gap may be
produced by exchanging two gluons in the $\hat t$ channel in a
color-singlet configuration (Fig.\ref{fig:c}b).
This is a $O(\a_s^4)$ process, but it is accompanied by infrared
logarithms due to the integration over the loop formed by the exchange of
the two gluons. Bjorken \cite{bj} estimates that
\beq
{\hsig_{sing}\over\hsig_{oct}} \sim 0.1\, .\label{diset}
\eeq
The probability that the gap is due to an electroweak exchange is rather small,
$\hsig_{\gamma,Z,W}/\hsig_{sing} \sim 10^{-2}$ \cite{bj,che},
and can be neglected. The radiation pattern for the emission of gluons
has been examined in detail in ref.~\cite{chez}. It has been found that
for parton-parton scattering with two-gluon exchange in a color-singlet
configuration the gluon radiation is suppressed in the rapidity interval
between the scattered partons, analogously to the suppression of gluon
radiation due to color coherence in photon exchange in the $\hat t$ channel.
Conversely, in one-gluon exchange the gluon radiation is found mainly in
the central rapidity region.

\subsection{The BFKL pomeron}
\label{sec:otto}

In the limit of high squared parton c.m. energy $\hat{s}$ and fixed $\hat{t}$,
we can describe the gap production
at the parton level by using the Balitsky-Fadin-Kuraev-Lipatov (BFKL) model
\cite{fkl,fklo,lip}, which resums the leading logarithmic contributions, in
$\ln(\hat{s}/\hat{t})$, to the scattering amplitudes to all orders in $\a_s$.
Therefore we may consider the exchange of
a two-gluon ladder in color-octet or -singlet configurations and compute
the leading $\ln(\hat{s}/\hat{t})$ virtual radiative corrections. For
a color-octet ladder we obtain \cite{fklo,mt,ddt},
\beq
{d\hsig_{oct}\over d\hat{t}} \simeq
{\pi N_c^2 \a_s^2\over 2\hat{t}^2}\, \exp\left(-{N_c\a_s\over\pi}\, \ln{\hat{s}
\over\hat{t}}\, \ln{\ptp^2\over\mu^2}\right)\, ,\label{diot}
\eeq
where $N_c=3$ is the number of colors, $\ptp$ is the transverse momentum
of the outgoing gluons, with $\hat{t} \simeq -\ptp^2$, and $\mu$ is a
cutoff which regulates the infrared divergence. Eq.(\ref{diot}) defines the
scattering as elastic if no soft gluons with $\ptp \gsim \mu$ appear in
the final state. The exponential of eq.(\ref{diot}) has the typical form of
a Sudakov form factor, and it vanishes as $\mu\ra 0$, in agreement with the
Bloch-Nordsieck behavior for bremsstrahlung emissions.
In addition, the exponential becomes smaller as the rapidity interval
between the partons
$\eta\simeq\ln(\hat{s}/\hat{t})$ grows, thus as we expected it is very
unlikely to produce a large gap between the partons through one-gluon exchange.

In the BFKL model the solution for the exchange of a color-singlet
two-gluon ladder is known only at $\hat{t} = 0$ \cite{fkl}, or at $\hat{t}
\ne 0$ for the scattering between colorless objects \cite{lip}.
Mueller and Tang \cite{mt} have modified the solution of ref.\cite{lip} in
order to describe the parton-parton elastic scattering at $\hat{t}\ne 0$.
Thus the elastic cross section for gluon-gluon scattering is \cite{mt,ddt},
\beq
{d\hsig_{sing}\over d\hat{t}} \simeq
{\pi^3 N_c^4 \a_s^4\over 4\hat{t}^2}\, {\exp\left(8\ln{2} {N_c\a_s\over\pi}\,
\ln{\hat{s}\over\hat{t}}\right)\over \left({7\over 2}\zeta(3) N_c\a_s
\ln{\hat{s}\over\hat{t}}\right)^3}\, .\label{dinov}
\eeq
Note that in the high-energy limit the singlet solution, eq.(\ref{dinov}),
does not depend on the infrared cutoff $\mu$ \cite{mt}. In addition, the
probability to produce a gap grows with the gap width, thus even though
higher-order the singlet solution quickly becomes more important than the
octet solution as the gap width grows.

Summing eq.(\ref{diot}) and (\ref{dinov}), and neglecting the ensuing double
counting which is not important at the large rapidities at which the BFKL
approximation applies, we obtain the probability of
producing a gap in gluon-gluon scattering. If $\mu\gg\lambda_{QCD}$ we obtain
the jet production rate with a gap in hadron production between the jets
by convoluting the sum of eq.(\ref{diot}) and (\ref{dinov}) with the parton
densities
\cite{mt,ddt}. This is legal as long as $\mu\gg\lambda_{QCD}$ because the
emission of soft hadrons in the rescattering between spectator partons in
the underlying event is allowed, in agreement with the factorization
theorems \cite{css}.

We compute the jet production rate as function of the
rapidity difference, $\Delta\eta=\eta_{j_1}-\eta_{j_2}$, and the rapidity
boost $\bar\eta$ (cf. sec.~\ref{sec:four}), since the elastic parton-parton
scattering does not depend on $\bar\eta$. Thus $\bar\eta$ may be fixed or
integrated out, thereby introducing a contribution due only to the variation
of the parton densities \cite{dds}. It is then convenient to compute the
gap fraction \cite{ddt}, i.e. to normalize the two-jet production with a
gap between the jets to the inclusive two-jet production\footnote{In
ref.~\cite{ddt,dds} the jets are ranked by
their rapidity, i.e. the two jets with the largest and smallest rapidity
are tagged, and the distribution is observed as a function of these two
tagging jets. Instead in the experiments \cite{gaps,bert,laura} the jets
are ranked by their transverse energies. However, there are preliminary
indications, at least in gap production in photoproduction events
\cite{laurel}, that the gap fraction does not change substantially ranking
the jets by their rapidity or by their transverse energies.}
\cite{dds} in order to minimize the normalization errors due to using the
BFKL approximation. Thus the gap fraction is,
\beq
\hat{f}(\mu\gg\lambda_{QCD}) = \hat{f}_{sing} + \hat{f}_{oct}\, ,\label{ventc}
\eeq
with $\hat{f}_{sing(oct)} = \hsig_{sing(oct)}/\hsig_{incl}$.
In ref.~\cite{ddt} the prediction for the
gap fraction at Tevatron energies as a function of the gap width
shows an abrupt rise of the gap fraction
at the largest gap widths kinematically allowed.
However much of it is not due to the growth of the singlet contribution in
the parton dynamics, eq.(\ref{dinov}), but merely to the parton luminosity,
which as $x\ra 1$ falls off faster for the inclusive two-jet production
than for the elastic one. This kinematic phenomenon is exactly the reverse
of the one noted for the $K$-factor in inclusive two-jet production in
ref.~\cite{dds,stir}. Therefore we conclude that at Tevatron energies within
the approximation of ref.~\cite{ddt} the gap fraction at large gap widths is
basically flat.

In comparing the prediction of ref.~\cite{ddt} with the experimental results
\cite{gaps,bert} a caveat is in order. The experiments measure the gap width,
$\Delta\eta_{gap}$, between the edges of the jet cones on the lego plot,
which differs from the rapidity difference, $\Delta\eta$, between the
outgoing partons originating the jets by the cone sizes $R$, i.e.
$\Delta\eta_{gap} = \Delta\eta - 2R$. The cone size used in the experiments
\cite{gaps,bert,laura} is $R = 0.7$. Within the BFKL approximation we cannot
distinguish between $\Delta\eta_{gap}$ and $\Delta\eta\simeq\ln(\hat{s}/
\hat{t})$ since in a leading logarithmic approximation the jets are
point-like.

In order to examine the gap fraction as a function of the gap width between
the jet-cone edges, while accounting properly for the cone structures, it is
necessary to perform a next-to-leading order calculation which includes
the basic features of color-singlet exchange. The simplest calculation of
this sort would be a full calculation of jet production at $O(\a_s^4)$.
At the moment this is in general unfeasible because one needs to know
the 4-parton 2-loop matrix elements, which have not been computed yet.
However the gap fraction, i.e. the ratio of the elastic to
the inclusive two-jet production, may be computed subtracting out from the
unity the ratio of the inelastic to the inclusive two-jet production, for
which at $O(\a_s^4)$ the 5-parton 1-loop matrix elements \cite{bdk}, and
the 6-parton tree-level matrix elements, suffice \cite{kos}.
The jet production with a gap in rapidity would then be computed by
requiring that any extra partons besides the ones we tag on be emitted
within the jet cones. Therefore a distinction between octet and singlet
contributions would not be done. In addition, the calculation would be
infrared stable.

\subsection{Gaps in soft-hadron production}
\label{sec:nove}

In sec.~\ref{sec:otto} we have required that the threshold $\mu$ in
soft-hadron production with respect to which we define the jet production as
elastic satisfy $\mu \gg \lambda_{QCD}$. Lowering the threshold to
$\mu \simeq \lambda_{QCD}$, as suggested by Bjorken \cite{bj} and
done in the experiments \cite{gaps,bert,laura}, the factorization
picture of ref.~\cite{css} does not apply, and we need a non-perturbative
model that lets the gap formed at the parton level survive
the rescattering between the spectator partons in the underlying event, which
would otherwise fill the gap with soft hadrons. Using an eikonal model,
Bjorken \cite{bj} estimates the rapidity-gap survival probability, $<|S^2|>$,
to be about 5-10\%. A study of several phenomenological models has also been
done in ref.~\cite{glm}.
In a first approximation we can then assume that the fraction of two-jet
events with a gap in soft-hadron production is \cite{bj},
\beq
f(\mu\ra\lambda_{QCD}) \simeq <|S^2|>\, \hat{f}(\mu\ra\lambda_{QCD})\,
,\label{ventun}
\eeq
with $\hat{f}$ as given in eq.(\ref{ventc}).
The survival probability, $<|S^2|>$, is expected to decrease as the
hadron-hadron c.m. energy $\sqrt{s}$ increases \cite{bj,glm}. Indeed
the total cross section, $\s_{tot}$, is related to the area of the
soft interactions, $\pi R^2$, and to the unitarity bound by the relation,
$\s_{tot} \simeq \pi R^2 \propto \ln{s^2}$. Thus as $s$ increases it is
less and less likely that the two hadrons do not interact.
Then $<|S^2|>$ is expected to be roughly independent of the gap width,
$\Delta\eta_{gap}$, since the rapidity interval between the jets $\Delta\eta$
is a kinematic parameter of the hard-interaction process, which according to
eq.(\ref{ventun}) would factor out of the soft interactions.
Finally, $<|S^2|>$ is expected to grow as the momentum fraction $x$ of the
incoming partons goes to 1, because there is less and less energy available
for the underlying event, i.e. for the spectator partons, in analogy with
the suppression of the underlying event observed in photoproduction events
as $x\ra 1$ \cite{jon}.

In gap production in photoproduction events at HERA, the smaller
c.m. energy $\sqrt{s}$, the smaller radius of the resolved photon
as compared to the proton, and the greater stiffness of the parton densities
in the photon as compared to the ones in the proton, all conspire
to make the value of $<|S^2|>$ larger. This might explain the higher value
of the gap fraction in the HERA data \cite{laura}, as compared to the one
in the Tevatron data \cite{gaps,bert}.

Since the theoretical calculation of $<|S^2|>$ is not very firm, it would be
better to measure it. In eq.(\ref{ventun}), $<|S^2|>$ appears
tangled to the gap fraction at the parton level, thus a single measurement
cannot give any information on $<|S^2|>$. However, if we raise the
threshold $\mu$ in such a way as to saturate the soft-gluon emission from the
underlying event as done in eq.(\ref{ventc}) then,
\beq
f(\mu\gg\lambda_{QCD}) \simeq \hat{f}(\mu\gg\lambda_{QCD})\, .\label{ventd}
\eeq
Subtracting the octet contribution \cite{bert}, the gap fraction
of eq.(\ref{ventun}) becomes,
\beq
(f-f_{oct})(\mu\ra\lambda_{QCD}) \simeq <|S^2|>\, (\hat{f}-\hat{f}_{oct})
(\mu\ra\lambda_{QCD})\, ,\label{ventf}
\eeq
Subtracting out the octet contribution from eq.(\ref{ventd}) and using the
insensitivity of the singlet contribution to the threshold $\mu$ we obtain,
\beq
(f-f_{oct})(\mu\gg\lambda_{QCD}) \simeq (\hat{f}-\hat{f}_{oct})
(\mu\gg\lambda_{QCD}) = (\hat{f}-\hat{f}_{oct})(\mu\ra\lambda_{QCD})\,
.\label{ventg}
\eeq
Thus taking the ratio of measurements of the gap fraction according to the
prescription of eq.(\ref{ventf}) and (\ref{ventg}) would allow us to
determine $<|S^2|>$.

\subsection{Conclusions}
\label{sec:ten}

The existence of rapidity gaps between jets in hadron-hadron \cite{gaps,bert}
and photon-hadron \cite{laura} collisions seems well established. The high
value of $|t|$ in these events suggests that short-distance strong-interaction
physics must play a vital role in them. Therefore these events may be
interpreted as evidence for a perturbative color-singlet exchange.

Bjorken's predictions, based on the simplest one-gluon and two-gluon
exchange and the survival of the rapidity gap at the soft-hadron level,
have been essentially confirmed by the data. Also the BFKL-pomeron picture
of ref.~\cite{mt,ddt}, which considers the radiative corrections to the
lowest-order one-gluon and two-gluon exchange, is in agreement with the data,
however the distinctive feature of the BFKL pomeron, i.e. the growth of the
gap fraction at very large gap widths, is not to be seen at the Tevatron.

There is large room for improvement of the model; on the theoretical side
we should make a more detailed model of color-singlet exchange at the parton
level, to keep into account the structure of the jets (cf.
sec.~\ref{sec:otto}); on the experimental side the
non-perturbative features of the gap production, like the rapidity-gap
survival probability (cf. sec.~\ref{sec:nove}), should be measured.

\section*{Acknowledgements}

I wish to thank bj Bjorken,
Andrew Brandt, Wilfried Buchm\"uller, Jon Butterworth,
John Collins, Yuri Dokshitzer, Dino Goulianos, Arthur Hebecker, Bob Hiroski,
Peppe Iacobucci, Brent May, Laurel Sinclair, Juan Terron, Jim Whitmore,
Mark W\"usthoff and Peter Zerwas for many useful discussions.
I would also like to acknowledge the hospitality of the CERN Theory Group
where part of this work was completed.

\end{document}